\documentclass[twocolumn,showpacs,preprintnumbers,
superscriptaddress]{revtex4}

\usepackage{amsfonts}
\usepackage{amsmath}
\usepackage{amssymb}
\usepackage{graphicx}

\begin{document}

\title{Theory of point contact spectroscopy in electron-doped cuprates}
\author{C. S. Liu}
\affiliation{Department of Physics, National Taiwan Normal University, Taipei 11650,
Taiwan}
\affiliation{Institute of Theoretical Physics, Chinese Academy of Sciences, Beijing
100080, China}
\author{W. C. Wu}
\affiliation{Department of Physics, National Taiwan Normal University, Taipei 11650,
Taiwan}
\pacs{74.20.-z, 74.45.+c, 74.50.+r, 74.25.Ha, 74.25.Jb.}

\begin{abstract}
In the hole-doped $d_{x^{2}-y^{2}}$-wave cuprate superconductor, due to the
midgap surface state (MSS), a zero bias conductance peak (ZBCP) is widely
observed in [110] interface point contact spectroscopy (PCS). However, ZBCP
of this geometry is rarely observed in the electron-doped cuprates, even
though their pairing symmetry is still likely the $d_{x^{2}-y^{2}}$-wave. We
argue that this is due to the coexistence of antiferromagnetic (AF) and the
superconducting (SC) orders. Generalizing the Blonder-Tinkham-Klapwijk (BTK)
formula to include an AF coupling, it is shown explicitly that the MSS is
destroyed by the AF order. The calculated PCS is in good agreement with the
experiments.
\end{abstract}

\maketitle


Pairing symmetry is an important issue towards the development of an
understanding the mechanism of superconductivity. For hole-doped high-$T_{c}$
cuprate superconductors, it is generally accepted that the pairing symmetry
is $d_{x^{2}-y^{2}}$-wave \cite{Tsuei969}. Among many supporting
experiments, PCS measurement shows a ZBCP due to the existing MSS in the
[110] direction 
\cite{Hu1526, Tanaka3451, deutscher:109}. On the electron-doped side of
cuprates, although no consensus has been reached yet, more and more recent
experiments have found results also consistent with a $d_{x^2-y^2}$-wave
pairing symmetry \cite{Sato2001, matsui017003, Armitage147003,
Ariando167001, blumberg107002, qazilbash:214510}. Thus one expected to see a
similar ZBCP for the electron-doped cuprates. The situation is more subtle
than this naive expectation however. The ZBCP has been consistently observed
in the underdoped samples. In the optimally- and over-doped samples, in
contrast, two doping-dependent coherence peaks were generally observed
\cite{Hayashi, Mourachkine, biswas207004, shan:144506}. This has been taken
to indicate that the excitation gap in electron-doped cuprates might switch
from $d$- to $s $-wave when the doping is increased -- a phenomenon
consistent with what was observed in the magnetic penetration depth
measurement \cite{kim087001}.

The clues towards an understanding of the complicated PCS of the
electron-doped cuprates lie in the doping evolution of the two (often called
$\alpha$ and $\beta$) pocket Fermi surfaces (FS) and the nonmonotonic $%
d_{x^{2}-y^{2}}$-wave excitation gap, as revealed by angle
resolved photoemission (ARPES) measurements \cite{Armitage257001,
matsui017003}. These phenomena have been interpreted in terms of a
phenomenological two-band model \cite{luo027001}, which in turn
led to a successful account for the magnetic penetration depth
measurement \cite{luo027001} and Raman scattering
\cite{liu:174517} data. The key feature of the two-band model is
that $\alpha $- and $\beta$-band superconducting (SC) gaps are
both monotonic $d_{x^2-y^2} $-wave, but with a different (doping
dependent) amplitude. Thus it was originally expected that [110]
MSS should also exist
in electron-doped cuprates. 
The difficulty in describing the experiments suggests that there may be some
other physics intervening in the system in the SC state.

The most promising candidate (for this absence) is the coexistence of an AF
order with the SC order. This scenario finds support in ARPES as the data
has been well explained in terms of a \textbf{k}-dependent band-folding
effect due to an existing AF order \cite{matsui047005, kusko140513, voo05,
yuan:054501, senechal:156404}. Apart from the ARPES data, no other direct
evidence indicates that AF order does indeed exist in these materials.
Therefore, it is highly desired that some other experiments provide more
definitive information regarding the possible existence of AF order in the
electron-doped cuprates. When a normal electron is incident into a
superconductor, it will induce the excitation of single quasi-particles
(QPs) corresponding to all possible ordering. The tunneling experiment is
thus considered to be one of the best choices. Among various tunneling
measurements, PCS is one of the most sensitive probes for the electronic
states.

In this paper, we shall focus on the effect of AF order on the PCS of a
normal metal-insulator-superconductor (NIS) junction. It will be shown
explicitly that the MSS can be destroyed by AF order. The observed PCS on
electron-doped cuprates is actually the result of competition between the
contributions from AF and SC orders.

Based on the simplest model that captures the essential physics,
we will first consider a superconductor overlayer which is coated
with a clean, size-quantized, normal-metal overlayer of thickness
$d$, that is much shorter than the mean free path $l$ of normal
electrons. The interface is assumed to be perfectly flat and
infinitely large. Considering $l\rightarrow \infty$, the
discontinuity of all parameters at the interface can be neglected,
except for the SC order parameter to which the proximity effect is
ignored \cite{Hu1526}. For $d_{x^2-y^2}$-wave superconductor of
interest, the interface is manipulated to be perpendicular to the
$\mathbf{k_{x}}$ axis (along the [110] direction). When both SC
and AF orders exist, QP excitations of an inhomogeneous
superconductor have a coupled electron-hole
character associated with the coupled $\mathbf{k}$ and $\mathbf{k+Q}$ [$%
\mathbf{Q}=(\pi ,\pi )$] subspaces. Correspondingly QP states can be
described by the generalized Bogoliubov-de Gennes (BdG) equations \cite%
{Gennes}
\begin{align}
Eu_{1}& =h_{0}u_{1}+\Delta _{\mathbf{k}}v_{1}+\Phi u_{2}  \notag \\
Ev_{1}& =\Delta _{\mathbf{k}}u_{1}-h_{0}v_{1}+\Phi v_{2}  \notag \\
Eu_{2}& =\Phi u_{1}+h_{0}u_{2}+\Delta _{\mathbf{k+Q}}v_{2}  \notag \\
Ev_{2}& =\Phi v_{1}+\Delta _{\mathbf{k+Q}}u_{2}-h_{0}v_{2},
\label{Bogoliubov-de Gennes equations}
\end{align}%
where $h_{0}\equiv -\hbar ^{2}\nabla _{\mathbf{x}}^{2}/2m-\mu $ with $\mu $
the chemical potential, $\Phi $ is the AF order parameter, and $\Delta _{\mathbf{k}}$ (%
$=\Delta _{0}\sin 2\theta $) is the $d_{x^2-y^2}$-wave SC order parameter
with $\Delta _{\mathbf{k+Q}}=-\Delta _{\mathbf{k}}$. Here the two-component
wave functions $u_{1}$ and $v_{1}$ are related to $\mathbf{k}$ subspaces,
while $u_{2}$ and $v_{2}$ are related to $\mathbf{k+Q}$ subspaces. In
arriving at Eq.~(\ref{Bogoliubov-de Gennes equations}), the pairing
potential is assumed to be $\sim\Delta _{\mathbf{k}}\Theta \left( x\right)$
with $\Theta \left( x\right) $ the Heaviside step function \cite{Hu1526} and
$\Delta _{\mathbf{k}}$, given above, the Fourier transform of the Cooper
pair order parameter in the relative coordinates.

\begin{figure}[ptb]
\begin{center}
\includegraphics[
width=7.5cm ]{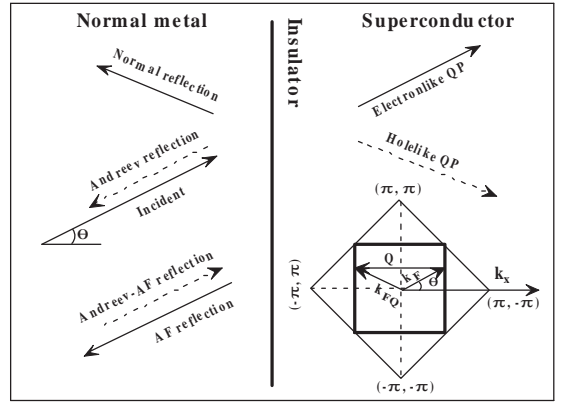}
\end{center}
\par
\vspace{-0.5cm} \caption{Schematic plot shows all possible
reflection and transmission processes for a normal electron incident
into an NIS junction, where an AF
order exists in the SC side. The inset shows the incident wave vector $%
\mathbf{k}_{F}=\left( k_{x0} ,k_{y},k_{z}\right) $ and its
corresponding AF wave vector
$\mathbf{k}_{F}+\mathbf{Q}\equiv\mathbf{k}_{F\mathbf{Q}}=\left(
-k_{x0},k_{y},k_{z}\right)$ due to an AF coupling. Both vectors are
tied to
the FS, which is approximated by a square (thick line). For convenience, $%
\mathbf{k_{x}}$ axis is chosen to be along the [110] direction.}
\label{fig1}
\end{figure}

To better describe the effect of an AF order, the FS will be approximated by
a square (see Fig.~\ref{fig1}). Thus at nearly half-filling, the FS matches
the MBZ boundary.
Under the WKBJ approximation \cite{Bardeen556} ($l=1,2$)
\begin{equation*}
\left(
\begin{array}{c}
u_{l} \\
v_{l}%
\end{array}%
\right) =\left(
\begin{array}{c}
{e}^{i\mathbf{k}_{F}\cdot \mathbf{r}}\tilde{u}_{l} \\
{e}^{-i\mathbf{k}_{F}\cdot \mathbf{r}}\tilde{v}_{l}%
\end{array}%
\right) ~\mathrm{and}~\left(
\begin{array}{c}
\tilde{u}_{l} \\
\tilde{v}_{l}%
\end{array}%
\right) =e^{-\gamma x}\left(
\begin{array}{c}
\hat{u}_{l} \\
\hat{v}_{l}%
\end{array}%
\right)
\end{equation*}%
with $\gamma $ being the attenuation constant for $\left\vert E\left(
\mathbf{k}_{F}\right) \right\vert <\left\vert \Delta \left( \mathbf{k}%
_{F}\right) \right\vert $ and $\mathbf{k}_{F}=\left(
k_{x},k_{y},k_{z}\right) $, Eq.~(\ref{Bogoliubov-de Gennes equations})
becomes the Andreev equation in the $\mathbf{k_{x}}$ direction
\begin{equation}
E\left(
\begin{array}{c}
\hat{u}_{1} \\
\hat{v}_{1} \\
\hat{u}_{2} \\
\hat{v}_{2}%
\end{array}%
\right) =\left(
\begin{array}{cccc}
\varepsilon & \Delta _{\mathbf{k}} & \Phi & 0 \\
\Delta _{\mathbf{k}} & -\varepsilon & 0 & \Phi \\
\Phi & 0 & -\varepsilon & -\Delta _{\mathbf{k}} \\
0 & \Phi & -\Delta _{\mathbf{k}} & \varepsilon%
\end{array}%
\right) \left(
\begin{array}{c}
\hat{u}_{1} \\
\hat{v}_{1} \\
\hat{u}_{2} \\
\hat{v}_{2}%
\end{array}%
\right)  \label{Andereev equations for SC}
\end{equation}%
for the superconducting overlayer ($x>0$). Here
$\varepsilon=\varepsilon(k_x)= i\gamma k_{x}/m$. The wave-vector
components parallel to the interface are conserved for all
possible processes.

Solving Eq.~(\ref{Andereev equations for SC}), one obtains eigenvalues $%
E=\pm \sqrt{\Delta _{\mathbf{k}}^{2}+\Phi ^{2}+\varepsilon ^{2}}$
where $+$ ($-$) corresponds to the electron- (hole-) like QP
excitation. Since $\varepsilon (-k_{x0})=-\varepsilon (k_{x0})$
and $\Delta \left( -k_{x0}\right) =-\Delta \left( k_{x0}\right)$,
states for $k_{x}=k_{x0}$ and $-k_{x0}$ are actually degenerate.
Thus for $k_{x}=k_{x0}$, one can have two degenerate eigenstates
for electron-like QP excitation, while for $k_{x}=-k_{x0}$, one
can have another two degenerate eigenstates for electron-like QP
excitation. Superposition of these four eigenstates thus gives a
formal wave function for the superconductor overlayer
\begin{gather}
\psi _{S}(x)=\left[ c_{1}\left(
\begin{array}{c}
\Delta _{\mathbf{k}} \\
E_{-} \\
0 \\
\Phi
\end{array}%
\right) +c_{2}\left(
\begin{array}{c}
E_{+} \\
\Delta _{\mathbf{k}} \\
\Phi  \\
0%
\end{array}%
\right) \right] e^{-\gamma x}e^{ik_{x0}x}  \notag \\
+\left[ c_{3}\left(
\begin{array}{c}
E_{-} \\
-\Delta _{\mathbf{k}} \\
\Phi  \\
0%
\end{array}%
\right) +c_{4}\left(
\begin{array}{c}
-\Delta _{\mathbf{k}} \\
E_{+} \\
0 \\
\Phi
\end{array}%
\right) \right] e^{-\gamma x}e^{-ik_{x0}x}.  \label{psi_s}
\end{gather}%
Here $E_{\pm }\equiv E\pm \varepsilon $ and $c_{i}$ are the coefficients of
corresponding waves. The wave function $\psi _{S}(x)$ remains the same if the
eigenstates of hole-like QP are made up. 


Solving Eq.~(\ref{Andereev equations for SC}) when both $\Delta _{\mathbf{k}%
} $ and $\Phi $ are set to be zero, one may obtain bound states to the
normal metal overlayer ($-d<x<0$). In this case, the eigenvalues become $%
E=\pm k_{x0}k_{1}/m$ assuming that the incident electron has the wave vector
$k_{x}=k_{1}$. At the interface, the wave functions of normal metal and
superconductor meet ideal continuity $\psi _{N}(x=0)=\psi _{S}(x=0)$. One
thus obtains the formal wave function for the normal metal overlayer
\begin{eqnarray}
&&\psi _{N}(x)=  \label{wave function for normal metal} \\
&&\left[ c_{1}\left(
\begin{array}{c}
e^{ik_{1}x}\Delta _{\mathbf{k}} \\
e^{-ik_{1}x}E_{-} \\
0 \\
e^{ik_{1}x}\Phi%
\end{array}%
\right) +c_{2}\left(
\begin{array}{c}
e^{ik_{1}x}E_{+} \\
e^{-ik_{1}x}\Delta _{\mathbf{k}} \\
e^{-ik_{1}x}\Phi \\
0%
\end{array}%
\right) \right] e^{ik_{x0}x}  \notag \\
&+&\left[ c_{3}\left(
\begin{array}{c}
e^{-ik_{1}x}E_{-} \\
-e^{ik_{1}x}\Delta _{\mathbf{k}} \\
e^{ik_{1}x}\Phi \\
0%
\end{array}%
\right) +c_{4}\left(
\begin{array}{c}
-e^{-ik_{1}x}\Delta _{\mathbf{k}} \\
e^{ik_{1}x}E_{+} \\
0 \\
e^{-ik_{1}x}\Phi%
\end{array}%
\right) \right] e^{-ik_{x0}x}.  \notag
\end{eqnarray}%
Considering the effect of free boundary at $x=-d$, it requires that $\psi
_{N}(x=-d)=0$. We then obtain the following eigencondition for the surface
bound states:
\begin{equation}
e^{-2ik_{1}d}E_{+}+e^{2ik_{1}d}E_{-}=2\Phi .  \label{eigencondition}
\end{equation}

Eq.~(\ref{eigencondition}) represents one of the major result in this paper.
When the AF order $\Phi =0$, there exists a zero-energy state which is
responsible for the ZBCP widely observed in hole-doped $d_{x^{2}-y^{2}}$%
-wave cuprate superconductors \cite{Hu1526}. When $\Phi \neq 0$, zero-energy
state no longer exists such that the energy of the existing state is always
finite. It is thus argued that the ZBCP does not exist in a system where the
AF and SC orders coexist, and it is the origin which led to the absence of
the ZBCP in the electron-doped cuprates. The above argument remains correct
when $d\rightarrow 0$, which corresponds to the case of no normal metal
overlayer (vacuum or with an insulating layer).


In order to compare the theory with experiments, we next consider a
superconductor [110] surface in point contact with a normal metal (STM) tip.
In this case, a thin insulating layer is considered to exist between the
normal metal and the superconductor (see Fig.~\ref{fig1}). For this kind of
NIS junction, the barrier potential is assumed to take a delta function, $%
V(x)=H\delta \left( x\right) $. Considering that an electron is injected
into the interface from the normal metal side (with an angle $\theta $),
four possible reflections and their related coefficients are detailed as
follows: (a) Normal reflection (reflected as electrons) with the coefficient
$r_{S}$. (b) Andreev reflection (reflected as holes, due to electron and
hole coupling in the $\mathbf{k}$ subspace) with the coefficient $r_{A}$.
(c) AF reflection (reflected as electrons, due to the $\mathbf{k}$ and $%
\mathbf{k}+\mathbf{Q}$ subspace coupling) with the coefficient $r_{AF}^{e}$.
(d) Andreev-AF reflection (reflected as holes, due to electron and hole
coupling in the $\mathbf{k}+\mathbf{Q}$ subspace) with the coefficient $%
r_{AF}^{h}$. In terms of these coefficients, the formal wave function for
the normal-metal side can then be written as
\begin{equation*}
\psi _{N}\left( x\right) =\left[
\begin{array}{c}
\exp \left( ik_{x0}x\right) +r_{S}\exp \left( -ik_{x0}x\right) \\
r_{A}\exp \left( ik_{x0}x\right) \\
r_{AF}^{e}\exp \left( ik_{x0}x\right) \\
r_{AF}^{h}\exp \left( -ik_{x0}x\right)%
\end{array}%
\right] .
\end{equation*}
The four reflection coefficients are determined by the boundary conditions:
\begin{align}
\psi _{N}\left( x\right) |_{x=0^{-}}& =\psi _{S}\left( x\right) |_{x=0^{+}},
\label{eq:boundary conditions} \\
\frac{2mH}{\hbar ^{2}}\psi _{S}\left( x\right) |_{x=0^{+}}& =\frac{d\psi
_{S}\left( x\right) }{dx}|_{x=0^{+}}-\frac{d\psi _{N}\left( x\right) }{dx}%
|_{x=0^{-}}.  \notag
\end{align}%
%
%
%
The normalized tunneling conductance is then given by
\begin{align}
\tilde{\sigma}\left( E,\theta \right) & =1-\left\vert r_{S}\left( E,\theta
\right) \right\vert ^{2}+\left\vert r_{A}\left( E,\theta \right) \right\vert
^{2}  \notag \\
& +\left\vert r_{AF}^{e}\left( E,\theta \right) \right\vert ^{2}-\left\vert
r_{AF}^{h}\left( E,\theta \right) \right\vert ^{2}.  \label{eq:sigma}
\end{align}


Actual tunneling conductance is intimately determined by the
junction properties. When the tip and the superconductor are in an
ideal point contact, \emph{i.e.}, when the effective potential
barrier $\hat{Z}\equiv 2mH/\hbar^{2}$ is sufficiently low and
narrow, the wave functions at the interface meet good condition of
continuity. Consequently Andreev and AF reflections are important.
For non-ideal point contacts, in contrast, only the normal
reflection is important. In the following, we focus on the case of
low and narrow barrier height, \emph{i.e.}, the PCS.

\begin{figure}[ptb]
\includegraphics[
width=9cm ]{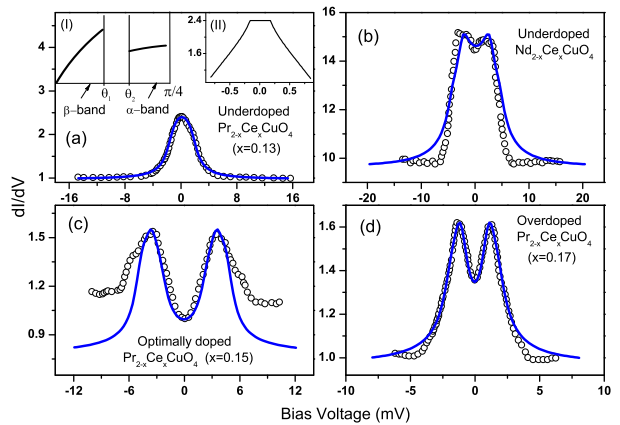} \vspace{-1.0cm} \caption{(Color online)
Fitting to the low-resistance $G$-$V$ data of electron-doped
cuprates. Data in frames (a),(c),(d) are taken from Ref.~\protect\cite%
{qazilbash024502} on Pr$_{2-x}$Ce$_{x}$CuO$_{4}$ at $T=1.43$K, while data in
frame (b) is taken from Ref.~\protect\cite{Mourachkine} on Nd$_{2-x}$Ce$_{x}$%
CuO$_{4}$. Fitting parameters are summarized in TABLE~\protect\ref{tab:table}%
. Inset (I) in (a) shows a typical piecewise $d_{x^{2}-y^{2}}$-wave gap,
with $\protect\theta_1$ ($\protect\theta_2$) characterizing the cutoff angle
for $\protect\beta$ ($\protect\alpha$) band FS. Inset (II) shows the $\hat{Z}%
\rightarrow 0$ calculation, emphasizing the zero-bias ``peak" structure in
(a).}
\label{fig2}
\end{figure}

Since the theory is aimed to the electron-doped cuprates, the issues
concerning their doping dependent AF and SC orders are crucial. Shown in the
inset (I) of Fig.~\ref{fig2}(a) is a typical SC gap for electron-doped
cuprates. The gap is piecewise: both segments are fitted to the monotonic $%
d_{x^2-y^2}$-wave ($\Delta_0 \sin 2\theta$), with different amplitudes ($%
\Delta_{0\beta}$ and $\Delta_{0\alpha}$). The $\beta$- and $\alpha$-band FS
segments are characterized by the cutoff angle $\theta_{1}$ and $\theta_{2}$%
. The values of doping-dependent $\theta_{1}$ and $\theta_{2}$ were
extracted from the ARPES data \cite{Armitage257001}.

In a real NIS junction experiment, the total tunneling conductance is given
by $\sigma(E)=\int \tilde{\sigma}\left(E,\theta\right)d\theta$. Here the
integration over the angle between $\theta_{1}$ and $\theta_{2}$ is ruled
out, to which the FS is absent. In the following the dimensionless
quantities $\Phi_\beta\equiv \Phi/\Delta_{0\beta}$ and $\Phi_\alpha\equiv
\Phi/\Delta_{0\alpha}$, which play the determining role on the actual PCS,
will be given from the fitting calculations. The actual values of $\Phi$, $%
\Delta_{0\alpha}$, and $\Delta_{0\beta}$ can also be obtained through the
fitting processes. Moreover in the calculation, the QP energy $E$ will be
replaced by $E-i\Gamma$ with $\Gamma$ characterizing the finite lifetime of
the QPs. The parameters from the best fittings are summarized in TABLE~\ref%
{tab:table}. As a matter of fact, only parameters associated with
the $\beta$ band ({\em i.e.}, associated with the nodal region)
are sensitive to the fitting.

\begin{table}[tbp]
\caption{Three (dimensionless) fitting parameters: $\Phi_\protect\beta$, $%
\Phi_\protect\alpha$, and $\hat{Z}$, obtained in Fig.~\protect\ref{fig2}. $%
\Delta_{0\protect\beta}$, $\Delta_{0\protect\alpha}$, and $\Phi$ (in unit of
meV) in the right three columns are determined exclusively from $\Phi_%
\protect\beta$ and $\Phi_\protect\alpha$ together with the actual peak
energy.}
\label{tab:table}%
\begin{ruledtabular}
\begin{tabular}{lccccccc}
 & $\Phi_\beta$ &$\Phi_\alpha$ & $\hat{Z}$
 &$ \Delta_{0\beta}$ & $\Delta_{0\alpha}$&$\Phi$ \\
\hline (a)& 0.10 & 0.20 & 0.05 & 1.88 & 0.94& 0.19
 \\
(c)& 0.75 & 1.00 & 1.50 & 3.00 & 2.25& 2.25
 \\
(d)& 0.70 & 0.75 & 0.75  & 1.33& 1.24&0.93
\\\hline (b)& 0.70 & 0.90 & 0.25  & 3.33 & 2.59& 2.33
 \\
\end{tabular}
\end{ruledtabular}
\end{table}

Figure~\ref{fig2}(a) compares the theoretical calculations with the
conductance-voltage ($G$-$V$) curve of PCS data \cite{qazilbash024502} on
underdoped Pr$_{2-x}$Ce$_{x}$CuO$_{4}$ (PCCO) ($x=0.13$). For this case, $%
\hat{Z}=0.05$ is used to simulate a low resistance $\left(R=9.8\Omega\right)$%
. The result is shown to be in good quantitative agreement with the data, to
which $(\Phi_{\beta},\Phi_{\alpha})=(0.1,0.2)$ are obtained. The smallness
of $\Phi_{\alpha}$ and $\Phi_{\beta}$ indicates that the effect of AF order
is inappreciable. As a consequence, a strong ZBCP appears, revealing the
dominance of the $d_{x^2-y^2}$-wave symmetry of the gap. This is in full
support to the conclusion drawn in Ref.~\cite{qazilbash024502} that the
paring symmetry for the underdoped samples is consistent with $d_{x^2-y^2}$%
-wave. To explore the AF effect in more details, the $G$-$V$ curve is
recalculated for an ideal point contact ($\hat{Z},\Gamma\rightarrow 0$). In
this limit [see inset (II) in Fig.~\ref{fig2}(a)], the zero-bias "peak" is
actually a mix with a narrow plateau. The plateau, that resembles the
feature of an $s$-wave superconductor, arises simply because a small but
finite AF order exists near the nodal region.

The evolution of the AF order can be understood from the
doping-dependent FS. In underdoped electron-doped cuprates, due to
their more distant FS from the MBZ, the scattering about
$\mathbf{Q}=(\pi,\pi)$ is weak. This results a small AF order.
With increasing the doping, FS approaching and crossing to the
MBZ, the scattering about $\mathbf{Q}$ becomes more important
which leads to a more important AF order \cite{yamada03,voo05}.
Upon further increase of the doping, long-range order is destroyed
such that the AF order decreases and vanishes eventually.

A similar plateau (with a dip) structure has also been observed in PCS on
underdoped Nd$_{2-x}$Ce$_{x}$CuO$_{4}$ \cite{Mourachkine}. To fit this set
of data, $(\Phi_{\beta},\Phi_{\alpha})=(0.7,0.9)$ and $\hat{Z}=0.25$ are
obtained [see Fig.~\ref{fig2}(b)]. Comparing the result with that in inset
(II) of Fig.~\ref{fig2}(a), the $G$-$V$ curve changes from a plateau to a
two-peak structure. The latter results from much higher $\Phi_{\beta}$ and $%
\Phi_{\alpha}$ and a slightly higher resistance $\hat{Z}$. The dip at zero
bias is an evidence that the ZBCP does not exist in this case.

Consider more closely how the FS segments evolve as the doping changes. At
low doping, $\alpha$-band FS first emerges in the antinodal direction before
the superconductivity sets in. When the doping is increased, $\beta$-band FS
appears simultaneously with the appearance of superconductivity. Since MSS
is the signature for a superconductor which has symmetrically a positive and
a negative portion of the gap, as long as $\Phi$ is inappreciable, the ZBCP
can still be observed in underdoped samples no matter how the FS segments
emerge, or even in the (unrealistic) case without the nodal ($\beta$-band)
FS.

PCS with high resistance $\hat{Z}$ has been observed on optimally-doped PCCO
($x=0.15$) \cite{qazilbash024502}. In the best fitting [Fig.~\ref{fig2}(c)],
large ratios $(\Phi_{\beta},\Phi_{\alpha})=(0.75,1.0)$ and $\hat{Z}=1.5$
(for $R=18\Omega$) are obtained. The strong effect of the AF order $\Phi$
leads to a clear two-peak feature, consistent with the case of a dominant $s$%
-wave gap. Nevertheless, theoretical curve deviating from the data at higher
biases confirms that a higher $\hat{Z}$ is in use and the wave functions at
the interface do not meet good condition of continuity.

In Fig.~\ref{fig2}(d), fairly good fitting is also made with the PCS data on
overdoped PCCO ($x=0.17$) \cite{qazilbash024502}, to which large $%
(\Phi_{\beta},\Phi_{\alpha})=(0.7,0.75)$ and a relatively smaller $\hat{Z}$
(for $R=2.6\Omega$) are used. Again, a two-peak feature consistent with an $%
s $-wave gap manifests the largeness of $\Phi$. In Ref.~\cite%
{qazilbash024502}, the $G$-$V$ curve in Fig.~\ref{fig2}(d) was also fitted
using the BTK model. Various pairing models were tested and it is the $d+is$%
-wave symmetry that leads to the best result. This does support the scenario
discussed in the current paper that single particle excitation is gapped by
both SC and AF orders, $\Delta_{\mathrm{eff}}\equiv|\Delta_d+i\Phi|=\sqrt{%
\Delta_d^{2}+\Phi^{2}}$. The actual PCS is indeed the competitive result
between the SC and AF contributions.


In summary, PCS of the electron-doped cuprates is investigated. It is shown
explicitly that MSS of the $d$-wave superconductor can be destroyed by the
presence of an AF order ($\Phi$). Due to the smallness of $\Phi$, ZBCP
occurs in the underdoped sample, consistent with the $d_{x^{2}-y^{2}}$-wave
pairing. A more important effect of $\Phi$ results a two-peak feature in
optimally and overdoped samples. The phenomenon that the peak energy first
increases and then decreases as doping increases is soundly explained.

We thank Profs. T. Xiang and H. G. Lou for useful comments. This work was
supported by National Science Council of Taiwan (Grant No.
94-2112-M-003-011) and National Natural Science Foundation of China (Grant
No. 10347149). We also acknowledge the support from the National Center for
Theoretical Sciences, Taiwan. 


\end{document}